\documentstyle[preprint,aps,eqsecnum,epsf]{revtex}                                       
\begin{document}                                                                
\preprint{UAHEP 9714}                                                            
\draft
\tightenlines
\title{Microfield Dynamics of Black Holes}                                      
\author{R. Casadio\thanks{                                                      
Permenent address: Dipartimento di Fisica dell'Universit\`a di Bologna          
and I.N.F.N., Sezione di Bologna, via Irnerio 46, I-40126 Bologna, Italy},      
B. Harms and Y. Leblanc}                                                        
\address{Department of Physics and Astronomy,                                   
The University of Alabama\\                                                     
Box 870324, Tuscaloosa, AL 35487-0324}                                          
\maketitle                                                                      
\begin{abstract}                                                                
The microcanonical treatment of black holes as opposed to the canonical         
formulation is reviewed and some major differences are displayed.               
In particular the decay rates are compared in the two different                 
pictures.                                                                       
\end{abstract}                                                                  
\section{Introduction}                                                          
In spite of its many mathematical and physical inconsistencies and              
drawbacks, the treatment of black holes as thermodynamical systems has          
since its inception been the description preferred by most                      
physicists investigating the nature of black holes.                             
Not least among the drawbacks is the fact that the laws of                      
quantum mechanics are violated, because the number density function of          
the emitted radiation as calculated using a thermal                             
vacuum is characteristic of mixed states, while the incoming radiation may      
have been in pure states.                                                       
Since black holes can in principle radiate away completely,                     
the unitarity principle is violated.                                            
\par                                                                            
In a series of papers                                                           
\cite{hl1,hl2,hl3,chl1,hl4,hl5,hl6,hl7,hl8,chlc,kndw}                           
we have investigated an alternative description of black holes which is         
free of the encountered problems in the thermodynamical approach.               
In our approach black holes are considered to be extended quantum objects       
($p$-branes).                                                                   
This point of view has recently been further supported by investigations        
in fundamental strings \cite{vafa}, where one finds that extended               
D-branes are a basic ingredient of the theory \cite{polchinski} and             
lead to black hole--type solutions for which the area of the horizon            
is proved to measure the quantum degeneracy \cite{maldacena}.                   
In the present work we consider a gas of $p$-brane black holes and show         
that the equilibrium configuration is decidedly non-thermal.                    
We also define a new vacuum, the microcanonical or fixed-energy vacuum,         
and obtain within the context of mean field theory the wave                     
functions for the radiation associated with such objects.                       
Using the number density function for our vacuum, we                            
calculate the black hole decay rate and compare it with that                    
obtained from the thermodynamical description.                                  
\par                                                                            
In Section~\ref{thermo} we present a brief summary of the thermodynamical       
description of processes involving black holes and discuss in detail the        
inconsistencies mentioned above.                                                
In Section~\ref{p-brane} we discuss our interpretation of                       
the WKB approximation as the quantum tunneling probability                      
and review our results for the statistical mechanics of a                       
gas of black holes.                                                             
In Section~\ref{mean} we discuss the thermodynamical interpretation of          
black holes within the context of mean field theory and prove that the          
thermal vacuum is the false vacuum for a black hole system.                     
We also present an alternative vacuum for such a system and the                 
microcanonical number density which corresponds to this vacuum.                 
In Section~\ref{wave} we present the microcanonical wave functions for          
the {\it in} and {\it out} states and in Section~\ref{decay} we derive          
the black hole decay rate.                                                      
\section{Thermodynamical Interpretation of Black Holes}                         
\label{thermo}                                                                  
Bekenstein's original observation \cite{bek} that the area of a black           
hole is in some way analogous to the thermodynamical concept of entropy         
was enlarged upon in Ref.\cite{bch} where the four laws of black hole           
thermodynamics were hypothesized.                                               
The mass difference of neighboring equilibrium states was shown to be           
related to the change in the black hole area $A$ according to the               
relation (Smarr formula)                                                        
\begin{eqnarray}                                                                
\Delta M = \kappa\,\Delta A +\varpi\,\Delta J+F\,\Delta Q                       
\; .                                                                            
\end{eqnarray}                                                                  
where $\kappa$ is the surface gravity and is related to                         
the temperature by                                                              
\begin{eqnarray}                                                                
T = \beta_H^{-1}={\kappa\over{2\pi}}                                            
\ .                                                                             
\label{betaH}                                                                   
\end{eqnarray}                                                                  
$J$ is the angular momentum of the black hole, $Q$ its charge and               
$\varpi$, $\Phi$ play the role of potentials.                                   
\par                                                                            
The partition function for the black hole is assumed to be                      
determined as                                                                   
\begin{eqnarray}                                                                
Z(\beta) = {\rm Tr}\, e^{-\beta\, H}= e^{-S_H}                                  
\; .                                                                            
\label{part}                                                                    
\end{eqnarray}                                                                  
The function $S_H$ is the Hawking entropy which is given by                     
\begin{eqnarray}                                                                
S_H &=&S_E- \beta_H\,\varpi\,J                                                  
\; ,                                                                            
\end{eqnarray}                                                                  
where $S_E$ is the Euclidean action.                                            
The Hawking entropy is also related to the area of the black hole by            
\begin{eqnarray}                                                                
S_H &=& {A\over{4}}                                                             
\; .                                                                            
\end{eqnarray}                                                                  
Finally, in thermodynamical equilibrium the statistical mechanical              
density of states is given by                                                   
\begin{eqnarray}                                                                
\Omega = Z^{-1}(\beta) = e^{S_H}                                                
\; ,                                                                            
\end{eqnarray}                                                                  
and the specific heat is                                                        
\begin{eqnarray}                                                                
C_V = {\partial E\over{\partial T}} = -{\beta^2\over{8\pi}}                     
\; .                                                                            
\end{eqnarray}                                                                  
The fact that the canonical specific heat, an intrinsically positive            
quantity, is negative in this interpretation is a clear signal that the         
thermodynamical analogy fails.                                                  
\par                                                                            
The thermodynamical interpretation of black holes has many such                 
inconsistencies.                                                                
A second problem can be best shown if we specialize the previous                
expressions, for instance, to the Schwarzschild black hole.                     
It turns out that                                                               
\begin{eqnarray}                                                                
S_H = S_E                                                                       
\; .                                                                            
\end{eqnarray}                                                                  
Further, since the radius of the horizon in this case is $2\,M$,                
one has                                                                         
\begin{eqnarray}                                                                
S_H = 4\,\pi\, M^2                                                              
\; ,                                                                            
\end{eqnarray}                                                                  
and                                                                             
\begin{eqnarray}                                                                
\beta_H = 8\,\pi\, M                                                            
\ .                                                                             
\end{eqnarray}                                                                  
It then follows that the partition function as calculated from the              
microcanonical density of states,                                               
\begin{eqnarray}                                                                
Z(\beta) &=& \int_0^\infty dM\, \Omega(M)\, e^{-\beta\, M}                      
 \nonumber \\                                                                   
&=& \int_0^\infty dM\, e^{4\,\pi\, M^2}\, e^{-\beta\, M} \to \infty             
\; ,                                                                            
\end{eqnarray}                                                                  
is infinite for all temperatures and hence the canonical ensemble               
is not equivalent to the (more fundamental) microcanonical ensemble             
\begin{eqnarray}                                                                
Z(\beta) \ne {1\over\Omega}                                                     
\; ,                                                                            
\end{eqnarray}                                                                  
as is required for thermodynamical equilibrium \cite{vanzo}                     
(see also Section~\ref{TFD-f} for a more general analysis).                     
\par                                                                            
Furthermore, if quantum mechanical effects are taken into account, black        
holes can be shown to radiate \cite{hawk,gibb}.                                 
In the thermodynamical approach the {\it out} vacuum is temperature             
dependent (see Section~\ref{wave}), and the radiation coming out of             
the black hole has a Planckian distribution                                     
\begin{eqnarray}                                                                
n_{\beta_H}(\omega)={1\over e^{\beta_H\,\omega}-1}                              
\ .                                                                             
\end{eqnarray}                                                                  
Since black holes can in principle radiate away completely,                     
this result implies that information can be lost, because                       
pure states can come into the black hole but only mixed states                  
come out.                                                                       
The breakdown of the unitarity principle is one of the most serious             
drawbacks of the thermodynamical interpretation, since it requires              
the replacement of quantum mechanics with some new (unspecified)                
physics.                                                                        
\section{Black Holes as P-Branes}                                               
\label{p-brane}                                                                 
The inconsistencies of the thermodynamical interpretation                       
are an indication that the interpretation of $e^{-S_H}$ as                      
the canonical partition function is wrong.                                      
In the usual WKB approximation $e^{-S_E}$ is the tunneling probability          
per unit volume for a particle to tunnel through a potential.                   
In the present case we hypothesize that the probability                         
to tunnel through the black hole's horizon is given by                          
\begin{eqnarray}                                                                
P \simeq e^{-S_H}                                                               
\; ,                                                                            
\end{eqnarray}                                                                  
for any kinds of black holes.                                                   
The quantum degeneracy of states for the system is proportional                 
to $P^{-1}$ and is then given by                                                
\begin{eqnarray}                                                                
\sigma \simeq c\,e^{A/4}                                                        
\ .                                                                             
\label{sig}                                                                     
\end{eqnarray}                                                                  
where the constant $c$ is determined from quantum field                         
theoretic corrections and can contain non-local effects.                        
Recently an analogous and maybe deeper understanding of Eq.~(\ref{sig}) has     
been obtained in string theory, where black hole solutions appear to be         
related to D-branes \cite{vafa,polchinski} and the relationship between area    
and entropy is recovered at least in the very special cases of tiny,            
extremal black holes \cite{maldacena}                                           
(the generalization to bigger, non extremal black holes might just be           
a technical problem, \cite{green}).                                             
\par                                                                            
Explicit expressions can be obtained for the above quantities for some          
geometries.                                                                     
\subsection{D-dimensional Schwarzschild black hole}                             
\label{schwa}                                                                   
As a first example we can consider the Schwarzschild black hole, which in       
$D$-dimensions has the Euclidean metric                                         
\begin{eqnarray}                                                                
ds^2 = e^{2\,\lambda}\, d\tau^2 + e^{-2\,\lambda}\, dr^2 + r^2\,                
d\Omega_{D-2}^2                                                                 
\; ,                                                                            
\end{eqnarray}                                                                  
where                                                                           
\begin{eqnarray}                                                                
e^{2\,\lambda} = 1 - \left({r_+\over{r}}\right)^{D-3}                           
\; .                                                                            
\end{eqnarray}                                                                  
The area in $D$ dimensions is                                                   
\begin{eqnarray}                                                                
{A\over{4}} = {A_{D-2}\over{16\,\pi}}\, \beta_H\, r_+^{D-3}                     
\; ,                                                                            
\end{eqnarray}                                                                  
with                                                                            
\begin{eqnarray}                                                                
M = {D-2\over{16\pi}}\, A_{D-2}\, r_+^{D-2}                                     
\; ,                                                                            
\end{eqnarray}                                                                  
where $A_{D-2}$ is the area of a unit $D-2$ sphere.                             
Eliminating the horizon radius $r_+$ in favor of the mass,                      
the area becomes                                                                
\begin{eqnarray}                                                                
{A\over{4}} = C(D)\, M^{D-2\over{D-3}}                                          
\; ,                                                                            
\end{eqnarray}                                                                  
where $C(D)$ is the mass-independent function                                   
\begin{eqnarray}                                                                
C(D) = {4^{D-1\over{D-3}}\, \pi^{D-2\over{D-3}} \over{(D-3)(D-                  
2)^{D-2\over{D-3}}\, A^{1\over{D-3}}_{D-2}}}                                    
\; .                                                                            
\end{eqnarray}                                                                  
Substituting in for $A/4$ in the degeneracy of states expression we find        
\begin{eqnarray}                                                                
\sigma(M) \simeq c\, e^{C(D)\,M^{D-2/D-3}}                                      
\; .                                                                            
\end{eqnarray}                                                                  
Comparing this expression to those known for non-local field                    
theories, we find that it corresponds to the degeneracy of                      
states for an extended quantum object ($p$-brane) of dimension                  
$p = {D-2\over{D-4}}$.                                                          
As has been demonstrated by several authors \cite{fub,deth,stru}, an            
exponentially rising density of states is the clear signal of a non-local       
field theory.                                                                   
$P$-brane theories are the only known non-local theories in                     
theoretical physics which can give rise to exponentially rising                 
degeneracies.                                                                   
\subsection{KND black hole}                                                     
In four dimensions the largest generalization of the Schwarzschild              
black hole is given by the Kerr-Newman family with the addition of              
a scalar field called the dilaton.                                              
The action of the Kerr-Newman dilaton (KND) black hole is found as              
an effective action in compactified string theories \cite{gsw}                  
and is given by                                                                 
\begin{equation}                                                                
S={1\over 16\,\pi}\int d^4x \,\sqrt{g}\,                                        
\left[R-{1\over 2}\,(\nabla\phi)^2-e^{-a\,\phi}\,F^2\right]+                    
\Sigma \ .                                                                      
\label{action}                                                                  
\end{equation}                                                                  
where the first term on the R.H.S. is the volume contribution                   
obtained by integrating on the whole region outside                             
the outer horizon, $R$ is the scalar curvature,                                 
$\phi$ is the dilaton field, $a$ its coupling constant,                         
$F$ is the Maxwell field and $\Sigma$ contains all surface terms.               
\par                                                                            
In Ref.~\cite{chlc} the field equations derived from the action in              
Eq.~(\ref{action}) were expanded in the charge-to-mass ratio,                   
$Q/M$, and a new perturbative static solution was found, which is               
of the form                                                                     
\begin{eqnarray}                                                                
ds^2 = -{\Delta\,\sin^2\theta\over\Psi}\,(dt)^2                                 
+ \Psi\,(d\varphi - \omega\,dt)^2                                               
+\rho^2\,\left[{(dr)^2\over\Delta}+(d\theta)^2\right]                           
\ .                                                                             
\label{g_ij}                                                                    
\end{eqnarray}                                                                  
The latter can be simplified upon substituting for the (bare)                   
parameters $M$, $Q$ and $J\equiv\alpha\,M$ the ADM mass, charge and             
angular momentum of the hole and also by shifting the radial coordinate,        
$r\to r-a^2\,Q^2/6\,M$ (see Ref.~\cite{kndw} for the details).                  
One finally obtains that the metric in Eq.~(\ref{g_ij}) coincides               
(at order $Q^2/M^2$) with the Kerr-Newman solution \cite{chan}.                 
This implies that the background dilaton field,                                 
\begin{eqnarray}                                                                
\phi=-a\,{r\over\rho^2}\,{Q^2\over M}+{\cal O}(Q^4)                             
\ ,                                                                             
\end{eqnarray}                                                                  
does not affect the singularity structure at order                              
$Q^2/M^2$.                                                                      
\par                                                                            
The surface term in Eq.~(\ref{action}) is given as                              
\begin{eqnarray}                                                                
\Sigma={\beta_H\over 2}\,M                                                      
\ ,                                                                             
\end{eqnarray}                                                                  
where the complexified time $i\,t$ has period $\beta_H$ as given in             
Eq.~(\ref{betaH}).                                                              
Also in this case there are two horizons,                                       
\begin{eqnarray}                                                                
r_\pm=M\pm\sqrt{M^2-\alpha^2- Q^2}                                              
\; ,                                                                            
\end{eqnarray}                                                                  
where $\alpha = J/M$ and the minimum value admitted for the mass is             
$M_0=\sqrt{\alpha^2+Q^2}$, corresponding to the {\em extremal\/} case.          
The Euclidean action of the KND instanton is                                    
\begin{eqnarray}                                                                
S_E(M,J,Q;a) = {A\over{4}} + \beta_H\, \varpi\, J                               
\; ,                                                                            
\end{eqnarray}                                                                  
and the area is given by                                                        
\begin{eqnarray}                                                                
A = 4\,\pi\,(r_+^2 + \alpha^2) +{\cal O}(Q^4)                                   
\; .                                                                            
\end{eqnarray}                                                                  
To order $Q^4$ the Euclidean action is                                          
\begin{eqnarray}                                                                
S_E(M,J,Q;a) = {\beta_H\over{2}}\,                                              
\left( M -{Q^2\,r_+ \over{r_+^2 + \alpha^2}}                                    
+ {\cal O}(Q^4)\right)                                                          
\; .                                                                            
\end{eqnarray}                                                                  
The quantum degeneracy of states is then                                        
\begin{eqnarray}                                                                
\sigma_{KND}(M,J,Q;a) \sim e^{A/4} = e^{\pi\,(r_+^2 + \alpha^2)}                
\; .                                                                            
\label{degen}                                                                   
\end{eqnarray}                                                                  
\subsection{Statistical mechanics of KND black holes}                           
Using the quantum degeneracy of states in Eq.~(\ref{degen}) we can analyze      
the statistical mechanical properties of a gas of such black holes.             
The microcanonical density is defined as a function in the space of all         
possible configurations of $n>0$ black holes                                    
\begin{eqnarray}                                                                
\Omega_{KND}(M,J,Q;a) = \sum_{n=1}^{\infty}\Omega_n(M,J,Q;a) \; .               
\label{omegn}                                                                   
\end{eqnarray}                                                                  
where $\Omega_n$ is determined from $\sigma_{KND}$ through                      
the relation                                                                    
\begin{eqnarray}                                                                
\Omega_n(M,J,Q;a)&=&\left[{V\over{(2\pi)}^3}\right]^n{1\over{n!}}\,             
\prod_{i=1}^n\, \int_{m_0}^\infty dm_i\,\int_{-m_i^2}^{+m_i^2}dj_i\,            
\int_{-\sqrt{m_i^2-\alpha_i^2}}^{+\sqrt{m_i^2-\alpha_i^2}} dq_i                 
\nonumber \\                                                                    
&&\times\sigma_{KND}(m_i,j_i,q_i;a)\,\int_{-\infty}^{+\infty} d^3p_i\,          
\delta\left(M-\sum\nolimits_{i=1}^n E_i\right)                                  
\nonumber \\                                                                    
&&\times\delta\left(Q-\sum\nolimits_{i=1}^n q_i\right)\,                        
\delta\left(J-\sum\nolimits_{i=1}^n j_i\right)\,                                
\delta^3\left(\sum\nolimits_{i=1}^n \vec{p}_i\right)                            
\; .                                                                            
\end{eqnarray}                                                                  
Here $M$, $J$ and $Q$ are respectively the total mass, angular momentum         
and charge of the gas, $m_0\ge 0$ is the minimum allowed mass for each          
black hole in the gas.                                                          
We are assuming in this relation that the black holes obey the                  
particle-like dispersion relation                                               
\begin{eqnarray}                                                                
E_i^2 = p_i^2 + m_i^2                                                           
\; .                                                                            
\end{eqnarray}                                                                  
The equilibrium configuration for such a gas is highly non-thermal.             
The most probable configuration turns out to be                                 
(see Ref.\cite{chl} for details) one massive black hole with                    
\begin{eqnarray}                                                                
&&{\rm Mass} = M-(n-1)\,m_0 \nonumber \\                                        
&&{\rm Charge} = Q - (n-1)\,\sqrt{1-\gamma^2}\,m_0 \\                           
&&{\rm Ang.\ Mom.} = J - (n-1)\,\gamma\, m_0^2,\ \ \ 0\le\gamma\le 1            
\ .                                                                             
\nonumber                                                                       
\end{eqnarray}                                                                  
The remaining $(n-1)$ black holes have                                          
\begin{eqnarray}                                                                
&&{\rm Mass} = m_0 \nonumber \\                                                 
&&{\rm Charge} = q_i =\sqrt{1-\gamma^2}\,m_0 \\                                 
&&{\rm Ang.\ Mom.} = j_i =\gamma\, m_0^2                                        
\ .                                                                             
\nonumber                                                                       
\end{eqnarray}                                                                  
A numerical computation was carried out for the special case $n =2$,            
$\alpha = 0$ (two Reissner-Nordstr\" om black holes).                           
As shown in Fig.1, the equipartition state is a saddle point, and     
the maxima correspond to one or the other of the                                
black holes possessing (nearly) all of the mass and all of                      
the charge.                                                                     
This suggests that                                                              
\begin{eqnarray}                                                                
\gamma \sim 0                                                                   
\; .                                                                            
\end{eqnarray}                                                                  
Our picture of the gas is thus one in which one massive black hole              
carries all the charge and angular momentum, and is surrounded by               
$n-1$ lighter, Schwarzschild black holes.                                       
Then the density of states can be approximated by                               
\begin{eqnarray}                                                                
\Omega(M,J,Q;a) &\sim& \left[{cV\over{(2\pi)^3}}                                
\right]^N\,{1\over{N!}}\, e^{(N-1)A_{KND}(m_0,\gamma m_0^2                      
,\sqrt{1-\gamma^2}m_0^2;a)/4}                                                   
\nonumber \\                                                                    
&&\phantom{\left[{cV\over{(2\pi)^3}}                                            
\right]^N\,}\times                                                              
e^{A_{KND}(M-(N-1)m_0, J-(N-1)m_0^2\gamma,Q-\sqrt{1-\gamma}(N-1)m_0;a)/4}       
\; .                                                                            
\end{eqnarray}                                                                  
This is the statistical mechanical model of a black hole                        
(the most massive one in the system) and its associated radiation               
(whose quanta are represented by the lighter black holes in the gas).           
%
%
%
%
\par                                                                            
The most probable number $N$ of black holes in the                              
gas is found from the extremum condition $d\Omega_n/dn|_{n=N}=0$.  The          
corresponding contribution                                                      
to the sum in Eq.~(\ref{omegn}) can be used to approximate                      
the full microcanonical density                                                 
\begin{equation}                                                                
\Omega(M,J,Q;V,a)\simeq \Omega_N(M,J,Q;V,a)                                     
\ .                                                                             
\label{omega}                                                                   
\end{equation}                                                                  
The number $N$ is given by $\Psi(N+1)\simeq\ln[c\,V/(2\,\pi)^3]$,               
where $\Psi$ is the psi function \cite{hl1}.                                    
\par                                                                            
We can now check whether the gas of black holes we have been describing         
satisfies the bootstrap condition \cite{hag},                                   
\begin{equation}                                                                
\lim\limits_{M\to\infty}\,{\Omega(M,Q,J;V,a)\over\sigma_{KND}(M,Q,J;a)}         
=1                                                                              
\ ,                                                                             
\label{boots}                                                                   
\end{equation}                                                                  
where $\sigma_{KND}$ is the quantum degeneracy of a single                      
black hole as given in Eq.~(\ref{degen}).                                       
For the general case in Eq.~(\ref{omega}), one obtains that                     
Eq.~(\ref{boots}) is satisfied if $m_0=0$ and                                   
\begin{equation}                                                                
\left[e^{\Psi(N+1)}\right]^N\,{1\over N!}=c                                     
\ .                                                                             
\end{equation}                                                                  
As in the case of a gas of Schwarzschild black holes \cite{hl1}, this           
equation gives a relation between the constant $c$ and the volume $V$.          
Correspondingly, one obtains the inverse microcanonical temperature             
\begin{eqnarray}                                                                
\beta ={d\ln\Omega \over d E}\simeq {d\ln\Omega_N\over dM}                      
=\beta_H(M,J,Q;a)                                                               
\ .                                                                             
\end{eqnarray}                                                                  
In the limit $a=0$ one recovers the Kerr-Newman expression,                     
$\beta_H=4\,\pi\,r_+$.                                                          
\par                                                                            
Our results show that the most probable state for a gas of                      
Kerr-Newman dilaton black holes is very far from thermal                        
equilibrium.                                                                    
Not only does one black hole acquire all of the mass                            
as in the Schwarzschild case, but it also acquires all of                       
the charge and all of the angular momentum of the whole gas.                    
This is the reason the bootstrap condition is satisfied for                     
the Kerr-Newman dilaton black holes at high energy.                             
\section{Mean Field Theory}                                                     
\label{mean}                                                                    
To study particle production and propagation in black hole geometries we        
now turn to a second semiclassical approximation.                               
In the mean field approximation fields are                                      
quantized on a classical black hole background.                                 
Black holes have non-trivial topologies which causally separate two             
regions of space.                                                               
For this reason the number of degrees of                                        
freedom is doubled, and two Fock spaces are required to                         
describe quantum processes occurring in the vicinity of a                       
black hole.                                                                     
Calculations of quantities associated with such                                 
processes can be carried out in ways analogous to calculations in               
Thermofield Dynamics (TFD) \cite{umez}, but with an overall fixed               
energy (Microfield Dynamics, or MFD, \cite{leblanc}).                           
\subsection{Canonical formulation}                                              
\label{TFD-f}                                                                   
In the context of mean field theory the thermal vacuum for quantum fields       
scattered off of black holes can be written as                                  
\begin{eqnarray}                                                                
|out;0\rangle  = {1\over{Z^{1/2}(\beta_H)}}\,\sum_{n=0}^{\infty}\,              
e^{-\beta_H\, n\,\omega/2}\,|n\rangle \otimes |\tilde{n}\rangle                 
\; ,                                                                            
\end{eqnarray}                                                                  
where the partition function $Z(\beta)$ is                                      
\begin{eqnarray}                                                                
Z = \sum_{n=0}^{\infty} e^{-\beta\, n\,\omega}                                  
\end{eqnarray}                                                                  
and the states $|\tilde{n}\rangle $ are a complete orthonormal basis for        
the region of space causally disconnected from an external observer.            
Operators corresponding to physically measurable quantities                     
are defined on the basis set $|n\rangle $ for states outside the                
horizon.                                                                        
The ensemble average (expectation value) of a physical observable               
$\hat O$ in the $out$ region is                                                 
\begin{eqnarray}                                                                
\langle out;0|{\hat O}|out;0\rangle  = {1\over{Z(\beta)}}\,                     
\sum_{n} e^{-n\,\beta\,\omega}\langle n|{\hat O}|n\rangle                       
\; .                                                                            
\label{expec}                                                                   
\end{eqnarray}                                                                  
As usual the temperature is determined by the surface gravity                   
according to Eq.~(\ref{betaH}).                                                 
\par                                                                            
For example if the operator $\hat O$ is the number operator                     
\begin{eqnarray}                                                                
{\hat O} = a^{\dagger}a                                                         
\; ,                                                                            
\end{eqnarray}                                                                  
for particles of rest mass $m$,                                                 
the ensemble average given in Eq.~(\ref{expec}) is the particle number          
density                                                                         
\begin{eqnarray}                                                                
n_{\beta_H}(m,k) = {1\over{e^{\beta_H\,\omega_k(m)}-1}}                         
\; .                                                                            
\label{n_k}                                                                     
\end{eqnarray}                                                                  
This expression can be immediately used for Schwarzschild black holes.          
Its generalization to particles (black holes) with spin and charge              
is straightforward and amounts to summing also over $J$ and $Q$                 
(which we avoid writing explicitly in this Section to keep the                  
notation simpler).                                                              
\par                                                                            
To describe particle interactions one needs the particle                        
propagator, which can be determined from Eq.~(\ref{expec})                      
by means of the time-ordered product $\langle out;0|T\phi(x_1)                  
\phi(x_2)|out;0\rangle $.                                                       
The Fourier transform of this expression is                                     
\begin{eqnarray}                                                                
\Delta_\beta = {1\over{k^2-m^2+i\epsilon}} - 2\,\pi\, i\,                       
\delta(k^2-m^2)\, n_\beta(m,k)                                                  
\; ,                                                                            
\label{delta}                                                                   
\end{eqnarray}                                                                  
with $n_\beta$ given by Eq.~(\ref{n_k}).  These expressions                     
are valid if black holes are described by a local field theory.                 
However, as discussed in Section~\ref{thermo}, the particle                     
number distribution given in Eq.~(\ref{n_k}) implies loss of                    
coherence.                                                                      
The $in$ state is a pure state                                                  
\begin{eqnarray}                                                                
|in;0\rangle  = |0\rangle  \otimes |\tilde{0}\rangle                            
\; ,                                                                            
\end{eqnarray}                                                                  
but the number density obtained from the outgoing states is                     
a thermal distribution.                                                         
\par                                                                            
In the microcanonical approach black holes are quantum excitations of           
$p$-branes, and hence non-local effects must be taken into account.             
This is accomplished by summing over all possible masses                        
(angular momenta and charges)                                                   
\begin{eqnarray}                                                                
n_{\beta_H}(k) = \int_0^{\infty}dm\, \sigma(m)\,n_{\beta_H}(m,k)                
\; .                                                                            
\end{eqnarray}                                                                  
Inclusion of non-local effects changes the thermal vacuum                       
to                                                                              
\begin{eqnarray}                                                                
|out;0\rangle  = {1\over{Z^{1/2}(\beta_H)}}                                     
\left[\prod_{m,k}\,\sum_{n_{k,m}=0}^{\infty}\right]\,\prod_{m,k}                
e^{-\beta_H\, n_{k,m}\omega_{k,m}/2}\, |n_{k,m}\rangle \otimes                  
|\tilde{n}_{k,m}\rangle                                                         
\;                                                                              
\end{eqnarray}                                                                  
in which the $m$ and $k$ labels are shown explicitly to emphasize the           
dependence of the states on the mass and momentum.                              
The quantity in square brackets represents the product of the sums over         
the discrete values of the momentum and mass.                                   
The canonical partition function extracted from this expression is              
\begin{eqnarray}                                                                
Z(\beta_H) = \exp\left(-{V\over{(2\pi)^{D-1}}}\int_{-\infty}^{+\infty}          
d^{D-1}{\vec k}\,\int_0^{\infty} dm \, \sigma(m)\,                              
\ln [1-e^{-\beta_H\,\omega_k(m)}]\right)                                        
\; ,                                                                            
\end{eqnarray}                                                                  
where the discrete mass and momentum indices have been                          
changed to continuous values.                                                   
A system in thermodynamical equilibrium must satisfy the condition              
\begin{eqnarray}                                                                
\int_0^{\infty}\Omega(E)\, e^{-\beta\,E_H}\, dE =                               
\exp\left(-{V\over{(2\,\pi)^{D-1}}}\,\int_{-\infty}^{+\infty}                   
d^{D-1}{\vec k}\, \int_0^{\infty}dm\, \sigma(m)\,                               
\ln [1-e^{-\beta_H\,\omega_k(m)}] \right)                                       
\; .                                                                            
\end{eqnarray}                                                                  
This is Hagedorn's self-consistency condition\cite{hag}.                        
It is well known that only strings ($p=1$) satisfy this condition               
\begin{eqnarray}                                                                
\sigma(m) \sim e^{b\; m} \; , \ \ \ \ (m \to \infty)                            
\; ,                                                                            
\end{eqnarray}                                                                  
for $\beta_H > b =$ Hagedorn's inverse temperature.                             
But black holes are not strings, as has been inferred from the                  
quantum mechanical density of states for Schwarzschild black holes              
in Section~\ref{schwa},                                                         
\begin{eqnarray}                                                                
\sigma(M) \sim e^{C\,M^{D-2/D-3}}                                               
\; .                                                                            
\end{eqnarray}                                                                  
\par                                                                            
Therefore black holes do not satisfy Hagedorn's condition                       
(${D-2\over D-3}>1$ for $D>3$).                                                 
The black hole system is not in thermal equilibrium because it does not         
fulfill the self-consistency condition required for thermal equilibrium.        
We are thus led to the conclusion that the thermal vacuum is the                
false vacuum for a black hole.                                                  
\subsection{Microcanonical formulation}                                         
The true vacuum for a black hole can be obtained by first                       
writing the thermal vacuum in terms of the density matrix                       
$\hat\rho$ for a system in thermal equilibrium                                  
\begin{eqnarray}                                                                
|0(\beta)\rangle  = \hat{\rho}(\beta ; {\bf H}) |\Im \rangle                    
\; ,                                                                            
\end{eqnarray}                                                                  
where                                                                           
\begin{eqnarray}                                                                
&&\hat{\rho}(\beta,{\bf H}) ={\rho(\beta,{\bf H})\over{\langle \Im|             
\rho(\beta,{\bf H})|\Im\rangle }}                                               
\nonumber \\                                                                    
&&{\rho} (\beta ; {\bf H}) = e^{-\beta {\bf H}}                                 
\nonumber \\                                                                    
&&|\Im\rangle  =                                                                
\left[\prod_{k,m}\sum_{n_{k,m}}\right]\,\prod_{k,m}|n_{k,m}\rangle              
\otimes |\tilde{n}_{k,m}\rangle                                                 
\; .                                                                            
\end{eqnarray}                                                                  
The trace of observable operators are given by                                  
\begin{eqnarray}                                                                
{\rm Tr}\,{\hat O} = \langle \Im | {\hat O}|\Im\rangle                          
\; .                                                                            
\end{eqnarray}                                                                  
For example the free field propagator can be determined from                    
\begin{eqnarray}                                                                
\Delta_{\beta}^{ab} = -i\langle \Im| T                                          
 \phi^a(x_1)\phi^b(x_2)\hat{\rho}|\Im\rangle                                    
\; .                                                                            
\end{eqnarray}                                                                  
The superscripts on $\phi$ refer to the member of the thermal                   
doublet \cite{umez}                                                             
\begin{eqnarray}                                                                
\phi^a = \left(\matrix{\phi \cr                                                 
\tilde{\phi}^{\dagger}\cr}\right)                                               
\;                                                                              
\end{eqnarray}                                                                  
being considered.                                                               
The Fourier transform of $\Delta_{\beta}^{11}(x_1,x_2)$                         
(the physical component) is equal to $\Delta_\beta$                             
given in Eq.~(\ref{delta}).                                                     
\par                                                                            
If instead of treating black holes as objects in                                
thermal equilibrium at fixed temperature $T$ and                                
corresponding vacuum $|\beta_H\rangle $ we treat the                            
black holes as having fixed energy $E$, we can                                  
formally define the microcanonical vacuum as                                    
\begin{eqnarray}                                                                
|E\rangle  = {1\over{\Omega(E)}}\int_0^E \Omega(E-E')\,                         
L_{E-E'}^{-1}[|\beta_H\rangle ]\,dE'                                            
\; ,                                                                            
\end{eqnarray}                                                                  
where $L^{-1}$ is the inverse Laplace transform.                                
Our analysis of the WKB approximation for black holes                           
in Sections~\ref{thermo} and~\ref{p-brane} shows that the assumption            
that black holes are at fixed energy is  physically more                        
reasonable than assuming that they are at fixed temperature.                    
Using this basis physical correlation functions are expressed as                
\begin{eqnarray}                                                                
G^{a_1,...,a_N}_{E}(1,2...,N) =                                                 
\langle \Im|T\phi^{a_1}(1),...,\phi^{a_N}(N)|E\rangle                           
\; .                                                                            
\end{eqnarray}                                                                  
\par                                                                            
Interaction effects can be taken into account by means of the                   
microcanonical propagator                                                       
\begin{eqnarray}                                                                
\Delta_E^{11}(k) =                                                              
{1\over{k^2-m^2+i\,\epsilon}} - 2\,\pi\, i\,\delta\,(k^2-m^2)\,                 
n_E(m,k)                                                                        
\; ,                                                                            
\end{eqnarray}                                                                  
where $n_E(m,k)$ is the microcanonical number density                           
\begin{eqnarray}                                                                
n_E(m,k) = \sum_{l=1}^\infty\,{\Omega(E-l\,\omega_k(m))                         
\over\Omega(E)}\,\theta(E-l\,\omega_k)                                          
\ ,                                                                             
\label{n_E}                                                                     
\end{eqnarray}                                                                  
which is our candidate alternative to Eq.~(\ref{n_k}) for the                   
distribution of particles emitted by a black hole.                              
\section{Wave functions}                                                        
\label{wave}                                                                    
The analysis carried on so far is global in nature.                             
In fact, although we were able to show consistent equilibrium                   
configurations for gases of black holes and number densities for                
the emitted radiation in such configurations,                                   
the geometry of spacetime never appears explicitly in the final                 
expressions.                                                                    
Of course, one is also interested in the local properties of                    
spacetime, and this is most intriguing in the present case                      
because the above results include implicitly any back-reactions                 
of the radiation on the metric.                                                 
\par                                                                            
Thus we need a probe to test the spacetime which corresponds to the             
microcanonical vacuum described in the previous Section.                        
We then turn to study the propagation of waves and show that the wave           
functions in the microcanonical vacuum can be obtained by making a              
formal replacement in the wave functions obtained for the thermal               
vacuum.                                                                         
\subsection{Thermal Vacuum}                                                     
In flat 4-dimensional spacetime with spherical coordinates                      
$t,r,\theta,\phi$ an incoming spherical wave collapsing                         
on a point is given in null coordinates asymptotically by                       
\begin{eqnarray}                                                                
\psi_{in} = {Y_{lm}(\theta,\phi)\over{\sqrt{8\,\pi^2\,\omega}}}\,               
{e^{-i\,\omega\, v}\over{r}}\ \ \ v = t+r_*                                     
\; ,                                                                            
\end{eqnarray}                                                                  
where $r_*$ is the so-called {\it tortoise} coordinate.                         
An outgoing wave has the form                                                   
\begin{eqnarray}                                                                
\psi_{out} = {Y_{lm}(\theta,\phi)\over{\sqrt{8\,\pi^2\,\omega}}}\,              
{e^{-i\,\omega\, u}\over{r}} \ \ \ u = t-r_*                                    
\; .                                                                            
\end{eqnarray}                                                                  
If we now consider waves propagating on a black hole                            
background, {\em e.g.} a Schwarzschild black hole, and do                       
not take into account back-reactions, then the incoming                         
wave becomes (see for example \cite{birr})                                      
\begin{eqnarray}                                                                
\psi_{in} = \left\{\matrix{\strut\displaystyle{ {Y_{lm}(\theta,\phi)            
\over{\sqrt{8\,\pi^2\,\omega}}}\,                                               
{e^{i\,(\omega/\kappa)\,\ln(v_0-v)}\over{r}}} &\ \ \ v < v_0\cr                 
\cr                                                                             
0 &\ \ \ v > v_0 \cr}\right.                                                    
\label{psi_T}                                                                   
\end{eqnarray}                                                                  
This wave obeys the wave equation in a background with                          
surface gravity $\kappa$.                                                       
The $in$ states for the two vacua are related by                                
the Bogolubov transformation                                                    
\begin{eqnarray}                                                                
\left.\matrix{\alpha_{\omega\omega'} \cr                                        
\cr                                                                             
\beta_{\omega\omega'}\cr}\right\} = {1\over{2\,\pi}}                            
\int_{-\infty}^{v_0} dv \,\left({\omega'\over{\omega}}\right)^{1/2}             
e^{\pm i\,\omega'\,v}\,e^{i\,(\omega/\kappa)\,\ln[(v_0-v)/c]}                   
\ ,                                                                             
\label{bogo}                                                                    
\end{eqnarray}                                                                  
where $c$ is a constant.                                                        
The two coefficients $\alpha$ and $\beta$ are related                           
by the Wronskian condition                                                      
\begin{eqnarray}                                                                
\sum_{\omega'}[|\alpha_{\omega\omega'}|^2 -                                     
|\beta_{\omega\omega'}|^2] = 1                                                  
\ .                                                                             
\label{omeg}                                                                    
\end{eqnarray}                                                                  
In Eq.~(\ref{omeg}) the variable $\omega'$ has been rendered                    
a discrete variable by box normalization of the wave functions.                 
Also backscattering of the fields from the                                      
spacetime curvature has been ignored.                                           
The integrals in Eq.~(\ref{bogo}) can be evaluated explicitly, and one          
finds that                                                                      
\begin{eqnarray}                                                                
|\alpha_{\omega\omega'}|^2 = e^{2\,\pi\,\omega/\kappa}\,                        
|\beta_{\omega\omega'}|^2                                                       
\ .                                                                             
\label{ab}                                                                      
\end{eqnarray}                                                                  
Substituting this relation into Eq.~(\ref{omeg}) one obtains                    
the Planckian distribution                                                      
\begin{eqnarray}                                                                
n_{\beta_H}(\omega) = \sum_{\omega'}|\beta_{\omega\omega'}|^2                   
= {1\over{e^{\beta_H\,\omega} - 1}}                                             
\; .                                                                            
\label{n_b}                                                                     
\end{eqnarray}                                                                  
\subsection{Microcanonical Vacuum}                                              
The relationship between $\alpha$ and $\beta$ in Eq.~(\ref{ab}) arises          
because the logarithmic term in Eq.~(\ref{bogo}) introduces a branch cut,       
and the integration around this branch cut causes the factor                    
multiplying this term (times $2\pi$) to appear in the                           
exponential multiplying $\beta$.                                                
Thus if we simply make the formal replacement                                   
\begin{eqnarray}                                                                
{2\,\pi\,\omega\over{\kappa}} \to \ln(1+n_E^{-1}(\omega))                       
\; ,                                                                            
\label{replace}                                                                 
\end{eqnarray}                                                                  
where $n_E(\omega)$ is the microcanonical number density                        
as expressed in Eq.~(\ref{n_E}), the relevant waves are of the form             
\begin{eqnarray}                                                                
\psi_{out} = {Y_{lm}(\theta,\phi)\over{\sqrt{8\,\pi^2\,\omega}}}\,              
{e^{-i\,\omega\, u}\over{r}}                                                    
\; ,                                                                            
\end{eqnarray}                                                                  
and                                                                             
\begin{eqnarray}                                                                
\psi_{in} = \left\{\matrix{\strut\displaystyle{ {Y_{lm}(\theta,\phi)            
\over{\sqrt{8\,\pi^2\,\omega}}}\,                                               
{e^{(i/2\,\pi)\,\ln[1+n_E^{-1}(\omega)]\,\ln(v_0-v)}\over{r}}}                  
&\ \ \ v < v_0\cr                                                               
\cr                                                                             
0 &\ \ \ v > v_0 \cr}\right.                                                    
\label{psi}                                                                     
\end{eqnarray}                                                                  
The relation between $\alpha$ and $\beta$ now becomes                           
\begin{eqnarray}                                                                
|\alpha_{\omega\omega'}|^2 = e ^{\ln(1+n_E^{-1}(\omega))}\,                     
|\beta_{\omega\omega'}|^2                                                       
\; ,                                                                            
\end{eqnarray}                                                                  
which gives for the sum over $\omega'$                                          
\begin{eqnarray}                                                                
\sum_{\omega'}|\beta_{\omega\omega'}|^2 = n_E(\omega)                           
\ .                                                                             
\end{eqnarray}                                                                  
Of course the wave in Eq.~(\ref{psi}) does not satisfy the same wave            
equation as the wave in Eq.~(\ref{psi_T}), but it will satisfy a wave           
equation in a background whose metric includes back-reaction and                
non-local effects.                                                              
\section{Black Hole Decay Rates}                                                
\label{decay}                                                                   
As an example of the differences between the predictions                        
of the two approaches (thermal vs. microcanonical)                              
suppose we consider 4-dimensional Schwarzschild black holes.                    
In this case the microcanonical density is given by                             
\begin{eqnarray}                                                                
\Omega(M) = e^{4\,\pi\, M^2}                                                    
\; ,                                                                            
\end{eqnarray}                                                                  
with the mass $M$ of the black hole being the fixed energy $E$.                 
Substituting the expression for $\Omega$ into Eq.~(\ref{n_E}),                  
we find for the number density                                                  
\begin{eqnarray}                                                                
n_M(\omega) = \sum_{l=1}^{M/\omega} e^{-8\,\pi\, M\,l\,\omega +                 
4\,\pi\, l^2\,\omega^2}                                                         
\; .                                                                            
\label{n_M}                                                                     
\end{eqnarray}                                                                  
If we ignore the $\omega^2$ term and let $M \to\infty$ at fixed $\omega$,       
we recover the thermal number density (Eq.~(\ref{n_b})).                        
This approximation works only for $\omega\ll M/l$, which shows that             
high frequency Hawking photons (with $\omega\sim M$) are responsible            
for the failure of the canonical description.                                   
Further, the $\omega^2$ term in Eq.~(\ref{n_M}) is a reflection of the          
fact that black holes have negative microcanonical specific heat.               
To see this let us rewrite Eq.~(\ref{n_E}) as                                   
\begin{eqnarray}                                                                
n_M &= &\sum_{l=1}^{\infty} {\Omega(M-l\,\omega_k(m))                           
\over{\Omega(M)}}\,\theta(M-l\,\omega_k)                                        
\nonumber\\                                                                     
&=& \sum_{l=1}^{M/\omega} \exp[S_E(M-l\,\omega) - S_E(M)]                       
\; .                                                                            
\end{eqnarray}                                                                  
Taylor-expanding the last expression we find                                    
\begin{eqnarray}                                                                
n_M = \exp\left(-l\,\omega\, {\partial S\over{\partial M}} -                    
{l^2\,\omega^2\over{2}}\, {\partial^2 S \over{\partial M^2}} + ...\right)       
\; .                                                                            
\end{eqnarray}                                                                  
The first term in the exponential is proportional to $\beta_H$                  
while the second term is proportional to $\beta^2_H/C_V$.                       
\par                                                                            
Using the expressions for $n_{\beta_H}$ in Eq.~(\ref{n_b}) and $n_M$            
in Eq.~(\ref{n_M}), we can compare the decay rates for radiating                
black holes predicted by the two theories.                                      
The luminosity is given by                                                      
\begin{eqnarray}                                                                
L = {1\over{2\pi}}\int_0^\infty d\omega \,\omega^3\,                            
\Gamma(\omega)\,n_i(\omega)                                                     
\ \ \ \ \ \ i = \beta, \,  M                                                    
\; .                                                                            
\end{eqnarray}                                                                  
The factor $\Gamma$ takes into account backscattering from                      
the spacetime curvature and in this capacity would replace the $1$ on           
the right hand side of Eq.~(\ref{omeg}).                                        
Carrying out the integrations for the number densities and multiplying          
by the horizon area to get the rate at which mass is lost, we find for          
the thermodynamical rate                                                        
\begin{eqnarray}                                                                
{dM\over{dt}} \sim -{1\over{M^2}}                                               
\; ,                                                                            
\end{eqnarray}                                                                  
and for the microcanonical rate                                                 
\begin{eqnarray}                                                                
{dM\over{dt}} \sim -M^6                                                         
\; .                                                                            
\end{eqnarray}                                                                  
The time dependences of the black hole mass are plotted                         
for these two cases in Figs.3 and 4.                                         
%
%
These graphs clearly indicate major differences in the decay rates              
predicted by the two theories.                                                  
The thermodynamical theory predicts that the decay rate blows up as             
$M \to 0$ and that it takes a finite time for the black hole to                 
completely evaporate,                                                           
\begin{eqnarray}                                                                
M\sim \left(M_0^3-3\,t\right)^{1/3}                                             
\ .                                                                             
\end{eqnarray}                                                                  
Thus one does not expect to find Fermi size ($10^{-13}$ cm) primordial          
black holes of mass less than $10^{15}$ gm in the present universe.             
The microcanonical rate, on the other hand, predicts that primordial,           
microscopic black holes could still be around today, since the                  
decay rate goes to zero as a power of $M$ and                                   
\begin{eqnarray}                                                                
M\sim {M_0\over\left(1+M_0^5\,t\right)^{1/5}}                                   
\ .                                                                             
\end{eqnarray}                                                                  
Of course both decay rates are unreliable when $M$ becomes of the               
order of the Planck mass and quantum gravity dominates.                         
\section{Conclusion}                                                            
The microcanonical approach is clearly preferable to the thermodynamical        
approach in the semiclassical quantization processes described above.           
It is free of the inconsistencies present in the thermodynamical approach,      
and its predictions seem to be more physically reasonable, {\em e.g.}           
a finite black hole decay rate through out the life of the black hole.          
The use of a fixed energy basis for the Hilbert space of the theory             
instead of the usual thermal state implies that black holes are particle        
states.                                                                         
In our interpretation of black holes as quantum objects the associated          
quantum degeneracy of states obtained from the inverse of the tunneling         
probability points to the identification of black holes with the excitation     
modes of $p$-branes.                                                            
The self-consistent treatment of black holes as quantum extended objects        
implies that black holes are elementary particles.                              
\section*{Acknowledgments}
This work was supported in part by the U.S. Department of Energy under Grant No. DE-FG02-96ER40967.

\begin{figure}
\caption{The total Bekenstein-Hawking entropy $S_{rn}$ for a system
of two Reissner-Nordstr\"om black holes with total mass $M=4$ and               
total charge $Q=1$ as a function of $m_1$ and $q_1$.}
\vspace{.25in}                           
\caption{When the lower limit for the mass of each black hole is                
$m_0=0.1$, the action $S_{rn}$ in Fig.1 has a maximum for             
$m_1=4-m_0$ and $q_1=1-m_0$, meaning that the second black hole                 
is extremal ($m_2=q_2=m_0$).}                                                   
\vspace{.25in}
\caption{The rate of black hole energy loss vs. time for the thermodynamical 
theory.\hspace{1in}}                                                               
\vspace{.25in}
\caption{The rate of black hole energy loss vs. time for the microcanonical
theory.\hspace{.6in}}                                                               
\end{figure}

\begin{references}                                                              
\bibitem{hl1}                                                                   
B. Harms and Y. Leblanc, {\it Phys. Rev. D} {\bf 46}, 2334 (1992).              
\bibitem{hl2}                                                                   
B. Harms and Y. Leblanc, {\it Phys. Rev. D} {\bf 47}, 2438 (1993).              
\bibitem{hl3}                                                                   
B. Harms and Y. Leblanc, {\it Ann. Phys.} {\bf 244},                            
262(1995); {\bf 244}, 272 (1995).                                               
\bibitem{chl1}                                                                  
P.H. Cox, B. Harms and Y. Leblanc, {\it Europhys. Letts.}                       
{\bf 26}, 321 (1994).                                                           
\bibitem{hl4}                                                                   
B. Harms and Y. Leblanc, {\it Europhys. Letts.} {\bf 27}, 557 (1994).           
\bibitem{hl5}                                                                   
B. Harms and Y. Leblanc, {\it Ann. Phys.} {\bf 242}, 265 (1995).                
\bibitem{hl6}                                                                   
B. Harms and Y. Leblanc,                                                        
{\it Proceedings of the Texas/PASCOS Conference, 92.                            
Relativistic Astrophysics and Particle Cosmology},                              
eds. C.W. Ackerlof and M.A. Srednicki,                                          
Annals of the New York  Academy of Sciences {\bf 688}, 454 (1993).              
\bibitem{hl7}                                                                   
B. Harms and Y. Leblanc,                                                        
{\it Supersymmetry and Unification of Fundamental Interactions},                
ed. Pran Nath, World Scientific (1994) p. 337.                                  
\bibitem{hl8}                                                                   
B. Harms and Y. Leblanc,                                                        
{\it Banff/CAP Workshop on Thermal Field Theory},                               
eds. F.C. Khanna, R. Kobes, G. Kunstatter and H. Umezawa,                       
World Scientific (1994), p. 387.                                                
\bibitem{chlc}                                                                  
R. Casadio, B. Harms, Y. Leblanc and P.H. Cox,                                  
{\it Phys. Rev. D} {\bf 55}, 814 (1997).                                        
\bibitem{kndw}                                                                  
R. Casadio, B. Harms, Y. Leblanc and P.H. Cox,                                  
{\it Phys. Rev. D} {\bf 56}, 4948 (1997).                                       
\bibitem{vafa}                                                                  
A. Strominger and C. Vafa, {\it Phys. Lett.} {\bf B379}, 99 (1996).             
\bibitem{polchinski}                                                            
J. Polchinski, {\it TASI Lectures on D-branes}, hep-th/9702136.                 
\bibitem{maldacena}                                                             
J. Maldacena, {\it Black holes and D-branes}, hep-th/9705078.                   
\bibitem{bek}                                                                   
J.D. Bekenstein, {\it Phys. Rev. D} {\bf 7} (1973) 2333;                        
{\it Phys. Rev. D} {\bf 9} (1974) 3292.                                         
\bibitem{bch}                                                                   
J.M. Bardeen, B. Carter, S.W. Hawking, {\it Commun. Math. Phys.} {\bf 31},      
161 (1973).                                                                     
\bibitem{vanzo}                                                                 
Black hole solutions whose Hawking entropy is such that a partition             
function can be consistently introduced have been recently found                
to exist in anti-de Sitter space, see L. Vanzo, {\it Black holes with           
unusual topology}, preprint gr-qc/9705004                                       
and private communication.                                                      
\bibitem{hawk}                                                                  
S.W. Hawking, {\it Comm. Math. Phys.} {\bf 43} (1975) 199.                      
\bibitem{gibb}                                                                  
G.W. Gibbons and S.W. Hawking {\it Phys. Rev. D}{\bf 15} (1977) 2752.           
\bibitem{green}                                                                 
M.B. Green, private communication.                                              
\bibitem{gsw}                                                                   
See, for example, M.B. Green, J.H. Schwarz and E. Witten,                       
{\it Superstring theory}, 2 Voll. Cambridge University Press,                   
Cambridge (1987).                                                               
\bibitem{fub}                                                                   
S. Fubini, A.J. Hanson and R. Jackiw, {\it Phys. Rev. D} {\bf7}, 1732 (1973).   
\bibitem{deth}                                                                  
J. Dethlefsen, H.B. Nielsen and H.C. Tze,                                       
{\it Phys. Lett.} {\bf 48B}, 48 (1974).                                         
\bibitem{stru}                                                                  
A. Strumia and G. Venturi, {\it Lett. Nuovo Cimento} {\bf 13}, 337 (1975).      
\bibitem{chan}                                                                  
S. Chandrasekhar,                                                               
{\it The Mathematical Theory of Black Holes},                                   
Oxford University Press, Oxford (1983).                                         
\bibitem{chl}                                                                   
R. Casadio, B. Harms and Y. Leblanc, {\em Statistical mechanics of              
Kerr-Newman dilaton black holes and the bootstrap condition}, preprint          
UAHEP 978, gr-qc/9706005, to be published in Phys. Rev. D.                      
\bibitem{hag}                                                                   
R. Hagedorn, {\it Nuovo Cimento Suppl.} {\bf 3}, 147 (1970);                    
S. Frautschi, {\it Phys. Rev. D} {\bf 3}, 2821 (1971);                          
R.D. Carlitz, {\it Phys. Rev. D} {\bf 5}, 3231 (1972).                          
\bibitem{umez}                                                                  
H. Umezawa, H. Matsumoto and M. Tachiki, {\it Thermo Field Dynamics             
and Condensed States}, North-Holland Publishing Co. Amsterdam, 1982.            
\bibitem{leblanc}                                                               
Y. Leblanc, in preparation.                                                     
\bibitem{birr}                                                                  
N.D. Birrell and P.C.W. Davies, {\it Quantum Fields in Curved Space},           
Cambridge University Press, Cambridge, 1982.                                    
\end{references}
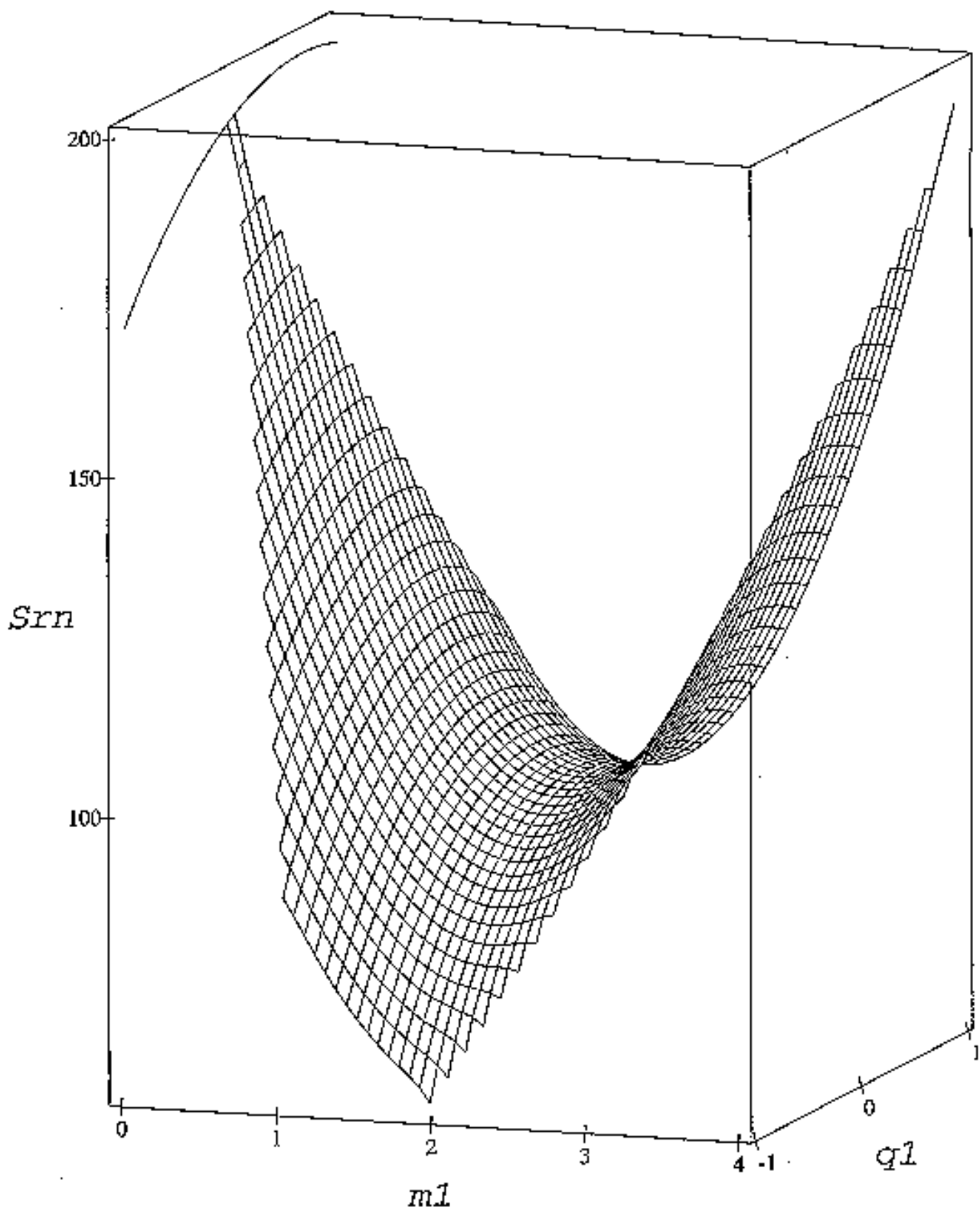
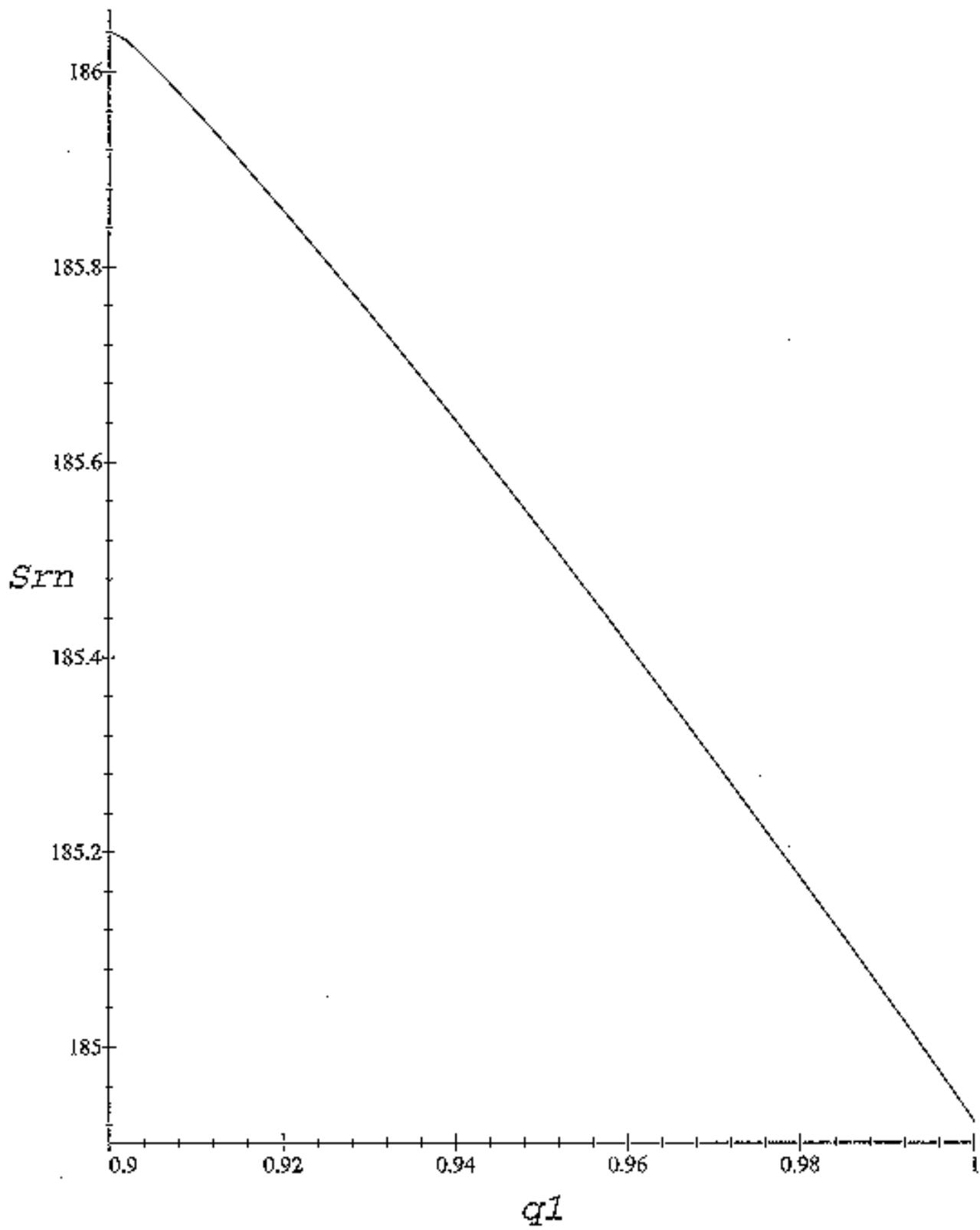
\end{document}